\documentclass{PoS}

\title{Non-renormalization theorem and cyclic Leibniz rule in lattice supersymmetry}

\ShortTitle{Non-renormalization theorem and cyclic Leibniz rule in lattice supersymmetry}

\author{Mitsuhiro Kato\\
 Institute of Physics, University of Tokyo, Komaba, Meguro-ku, Tokyo 153-8902, Japan \\
        E-mail: \email{kato@hep1.c.u-tokyo.ac.jp}}
\author{\speaker{Makoto Sakamoto}%
         \thanks{
         This work was supported in part by
         Grants-in-Aid for Scientific Research (No.25400260 and No.25287049) 
         by the Japanese Ministry of Education, Science, Sports and Culture.}\\
         Department of Physics, Kobe University, Nada-ku, Hyogo 657-8501, Japan\\
         E-mail: \email{dragon@kobe-u.ac.jp}}
\author{Hiroto So\\
    Department of Physics,  Ehime University, Bunkyou-chou 2-5, 
 Matsuyama 790-8577, Japan\\
        E-mail: \email{so@phys.sci.ehime-u.ac.jp}}

\abstract{
We propose a lattice model of supersymmetric complex quantum
mechanics which realizes the non-renormalization theorem
on a lattice.
In our lattice model, the Leibniz rule in the continuum,
which cannot hold on a lattice due to a no-go theorem, 
is replaced by the cyclic Leibniz rule (CLR) for 
difference operators.
It is shown that CLR allows two of four supercharges
of the continuum theory to preserve while a naive
lattice model can realize one supercharge at the most.
A striking feature of our lattice model is that there
are no quantum corrections to potential terms in any finite
order of perturbation theory.
This is one of characteristic properties of supersymmetric theories
in the continuum.
It turns out that CLR plays a crucial role in the
proof of the non-renormalization theorem.
This result suggests that CLR grasps an essence of
supersymmetry on a lattice.
}
\FullConference{The 32nd International Symposium on Lattice Field Theory,\\
		23-28 June, 2014\\
		Columbia University New York, NY}

\begin{document}
%
%

%
\section{Introduction}
%
%
For more than thirty years, no one has succeeded to construct 
satisfactory lattice models which realize a full supersymmetry
algebra.
One might try to find lattice supersymmetry transformations 
$\delta_{Q}, \delta_{Q'}$ such that a lattice action is invariant
under $\delta_{Q}, \delta_{Q'}$ with a supersymmetry algebra
$\{ \delta_{Q} , \delta_{Q'} \} = \delta_{P}$, where $\delta_{P}$
would correspond to some \lq\lq translation\rq\rq\ on a lattice.
It, however, turns out to be hard to realize it.
If we expect $\delta_{P}$ to act on any field $\phi$ as
$\delta_{P}\phi=\Delta\phi$ with some difference operator
$\Delta$ on a lattice, it may contradict with a no-go theorem
on the Leibniz rule \cite{no-go}:
The $\delta_{P}$ is assumed to satisfy the Leibniz rule
$\delta_{P}(\phi\rho) = (\delta_{P} \phi)\rho+\phi(\delta_{P}\rho)$,
but any difference operator $\Delta$ cannot obey the rule
under some reasonable assumptions due to the no-go theorem.

%
One of interesting approaches to construct supersymmetric lattice 
models is to use Nicolai maps \cite{sakai-sakamoto}, which are known as
a characterization of supersymmetric theories \cite{nicolai}.
It has been observed that a Nicolai map realizes a nilpotent
supersymmetry \cite{NicolaiMap2}, which avoids the problem of a
lack of a Leibniz rule on a lattice.
However, since such a nilpotent supersymmetry can be defined without
using difference operators, it is doubtful whether the nilpotent
supersymmetry by itself is enough
to keep crucial features of supersymmetry, e.g. the non-renormalization
of superpotentials \cite{nonrenormalization}.

%
In this paper, we investigate supersymmetric complex quantum mechanics
on a lattice.
We propose a cyclic Leibniz rule, in place of the Leibniz rule,
which will be found to be a useful notion in lattice supersymmetry,
and show that our lattice model can possess an $N=2$ nilpotent
supersymmetry algebra with the help of the cyclic Leibniz rule.
A striking feature of our lattice supersymmetric model is that
a non-renormalization theorem holds for a finite lattice spacing,
even though a full supersymmetry algebra is not realized in our
lattice model.

%
\section{Leibniz rule vs. cyclic Leibniz rule}
%
%
In order to explain the difference between the Leibniz rule and 
the cyclic Leibniz rule, we start with a setup of our lattice model.
A difference operator $\Delta$ on a field $\phi_{n}$, a field 
product $(\phi * \rho)_{n}$ and an inner product $\langle\phi,\rho\rangle$ 
between $\phi_{n}$ and $\rho_{n}$ are defined as
%
\begin{eqnarray}
(\Delta\phi)_{n} \equiv \sum_{m} \Delta_{nm}\,\phi_{m}\,, \quad
(\phi * \rho)_{n} \equiv \sum_{ml}M_{nml}\,\phi_{m}\,\rho_{l}\,, \quad
\langle\phi\,,\,\rho\rangle \equiv \sum_{n}\phi_{n}\,\rho_{n}\,,
\label{setup}
\end{eqnarray}
%
where the indices $l, m, n$ denote positions on a lattice.
We require that the coefficients $\Delta_{nm}$ and $M_{nml}$ are
translationally invariant, such that $\Delta_{nm}=\Delta(n-m)$ and
$M_{nml}=M(n-m,n-l)$.
We further require the difference operator and the field product to be local.
The locality property is important in constructing local field theories
after the continuum limit.
It is sufficient for our purposes to restrict the field product 
$\phi * \rho$ to be symmetric for
any bosonic fields $\phi$ and $\rho$, i.e.
$\phi * \rho = \rho * \phi$.

In terms of the three-body product 
$\langle\phi^{(1)}\,,\,\phi^{(2)}*\phi^{(3)}\rangle$,
the Leibniz rule would be represented as
%
\begin{eqnarray}
\langle\Delta\phi^{(1)}\,,\,\phi^{(2)}*\phi^{(3)}\rangle
 + \langle\phi^{(1)}\,,\,\Delta\phi^{(2)}*\phi^{(3)}\rangle
  + \langle\phi^{(1)}\,,\,\phi^{(2)}*\Delta\phi^{(3)}\rangle = 0.
\label{LR}
\end{eqnarray}
%
An important observation is that any difference operators and field
products cannot satisfy the relation (\ref{LR}) with the properties
(i) the locality and (ii) the translation invariance \cite{no-go}.
Therefore, we have to abandon the realization of the Leibniz rule on a lattice.

Instead of the Leibniz rule, we here propose the cyclic Leibniz rule, which is a 
modified version of the Leibniz rule, defined by
%
\begin{eqnarray}
\langle\Delta\phi^{(1)}\,,\,\phi^{(2)}*\phi^{(3)}\rangle
 + \langle\Delta\phi^{(2)}\,,\,\phi^{(3)}*\phi^{(1)}\rangle
  + \langle\Delta\phi^{(3)}\,,\,\phi^{(1)}*\phi^{(2)}\rangle = 0
\label{CLR}
\end{eqnarray}
%
for any bosonic fields $\phi^{(i)}$ $(i=1,2,3)$.
It is interesting to note that by taking all $\phi^{(i)}$ $(i=1,2,3)$ to be
equal the cyclic Leibniz rule (\ref{CLR}) reduces to the relation
%
\begin{eqnarray}
\langle\Delta\phi\,,\,\phi*\phi\rangle = 0.
\label{totalderivative}
\end{eqnarray}
%
This is an analog of a vanishing total divergence 
$\int dx\, (\partial_{x}\phi)\phi^{2} = \frac{1}{3}\int dx\, \partial_{x}(\phi^{3}) = 0$ 
in the continuum.
The relation (\ref{totalderivative}) will play an important role in constructing
lattice supersymmetric models because Lagrangians are not invariant, in general, 
but become total divergences under supersymmetry transformations.

A difference between the Leibniz rule and the cyclic one is the 
position of the difference operator.
The difference operator $\Delta$ always acts on the leftmost field
in each term of the cyclic Leibniz rule (\ref{CLR}).
Another difference is the order of the fields: $\phi^{(i)} (i=1,2,3)$
are placed in the same order in every term of eq.(\ref{LR}), 
while they are cyclically permutated in eq.(\ref{CLR}).
This is the reason why we call the relation (\ref{CLR}) a cyclic Leibniz rule.
The extension of eq.(\ref{CLR}) to $k$-body products will be obvious:
%
\begin{eqnarray}
&&\langle\Delta\phi^{(1)}\,,\,\phi^{(2)}*\phi^{(3)}*\cdots *\phi^{(k)}\rangle
 + \langle\Delta\phi^{(2)}\,,\,\phi^{(3)}*\cdots *\phi^{(k)}*\phi^{(1)}\rangle
  \nonumber\\
  &&\qquad + \cdots
   + \langle\Delta\phi^{(k)}\,,\,\phi^{(1)}*\phi^{(2)}*\cdots *\phi^{(k-1)}\rangle = 0\,.
\label{k-CLR}
\end{eqnarray}
%

%

%
We should emphasize that the cyclic Leibniz rule (\ref{CLR})
can be realized on a lattice with the properties of the locality and the 
translation invariance.
In fact, the following difference operator and field product
(with a unit lattice spacing)
%
\begin{eqnarray}
(\Delta\phi)_{n}
  &=& \frac{1}{2}(\phi_{n+1} - \phi_{n-1})\,,\nonumber\\
(\phi*\rho)_{n}
  &=& \frac{1}{2}(2\phi_{n+1}\rho_{n+1} + 2\phi_{n-1}\rho_{n-1}
                  +\phi_{n+1}\rho_{n-1} + \phi_{n-1}\rho_{n+1})
\label{example_CLR}
\end{eqnarray}
%
are shown to satisfy (i) the locality, (ii) the translation invariance
and (iii) the cyclic Leibniz rule.
General forms of difference operators and field products satisfying
the cyclic Leibniz rule have been studied in Ref.\cite{CLR}.
We would like to note that the trivial field product
$(\phi*\rho)_{n} = \phi_{n}\rho_{n}$,
which is usually used in the literature,
cannot satisfy the cyclic Leibniz rule for any difference operators.
This is because the cyclic Leibniz rule (\ref{CLR})  for the trivial field product
reduces to the Leibniz rule (\ref{LR}), which cannot be realized by any
difference operators.
Therefore, the field product $\phi*\rho$ has to be nontrivial in order to
obtain the cyclic Leibniz rule (\ref{CLR}) on a lattice.

%
\section{Supersymmetric complex quantum mechanics on a lattice}
%
%
In this paper, we consider the following lattice action of 
supersymmetric complex quantum mechanics \cite{Lattice2013}:
%
\begin{eqnarray}
S &=& \langle\Delta\phi_{-}\,,\,\Delta\phi_{+}\rangle
    - \langle F_{-}\,,\,F_{+}\rangle
    - i\langle\chi_{-}\,,\,\Delta\bar{\chi}_{+}\rangle
    + i\langle\Delta\bar{\chi}_{-}\,,\,\chi_{+}\rangle \nonumber\\
  &&- m_{+} \langle F_{+}\,,\,G\phi_{+}\rangle
    + m_{+}\langle\chi_{+}\,,\,G\bar{\chi}_{+}\rangle
    - \lambda_{+}\langle F_{+}\,,\,\phi_{+}*\phi_{+}\rangle
    + 2\lambda_{+}\langle\chi_{+}\,,\,\bar{\chi}_{+}*\phi_{+}\rangle \nonumber\\
  &&- m_{-} \langle F_{-}\,,\,G\phi_{-}\rangle
    - m_{-}\langle\chi_{-}\,,\,G\bar{\chi}_{-}\rangle
    - \lambda_{-}\langle F_{-}\,,\,\phi_{-}*\phi_{-}\rangle
    - 2\lambda_{-}\langle\chi_{-}\,,\,\bar{\chi}_{-}*\phi_{-}\rangle\,,
\label{SCQM}
\end{eqnarray}
%
where $\phi_{\pm}$ and $F_{\pm}$ ($\chi_{\pm}$ and $\bar{\chi}_{\pm}$)
are complex bosonic (fermionic) variables.
The terms proportional to $m_{\pm}$ correspond to the mass terms for
$\phi_{\pm}, \chi_{\pm}, \bar{\chi}_{\pm}$ and they can also include
Wilson terms to avoid fermion doublings.\footnote{%
By taking $(G\phi)_{n}=\sum_{m}(\delta_{nm}+A_{nm})\phi_{m}$ in eq.(\ref{SCQM}),
we can introduce a Wilson term by use of the coefficient $A_{nm}$ as well as
a mass term in the first term \cite{CLR}.
}

We require that the lattice action (\ref{SCQM}) is exactly invariant under 
the $N=2$  nilpotent supersymmetry transformations
%
\begin{eqnarray}
\left\{
 \begin{array}{l}
   \delta_{+}\phi_{+}=\bar{\chi}_{+}\,,\\
   \delta_{+}\chi_{+}=F_{+}\,,\\   
   \delta_{+}\chi_{-}=-i\Delta\phi_{-}\,,\\
   \delta_{+}F_{-}=-i\Delta\bar{\chi}_{-}\,,\\
   \textrm{others}=0\,,\\
 \end{array} \right.
\qquad\qquad
\left\{
 \begin{array}{l}
   \delta_{-}\chi_{+}=i\Delta\phi_{+}\,,\\
   \delta_{-}F_{+}=-i\Delta\bar{\chi}_{+}\,,\\   
   \delta_{-}\phi_{-}=-\bar{\chi}_{-}\,,\\
   \delta_{-}\chi_{-}=F_{-}\,,\\
   \textrm{others}=0\,,\\
 \end{array} \right.
\label{SUSYtransformation}
\end{eqnarray}
%
which satisfy the $N=2$ nilpotent supersymmetry algebra
%
\begin{eqnarray}
(\delta_{+})^{2} = (\delta_{-})^{2} = \{\,\delta_{+}\,,\,\delta_{-}\,\} = 0\,.
\label{nilpotentalgebra}
\end{eqnarray}
%
The requirement that the lattice action (\ref{SCQM}) is invariant
under the transformations (\ref{SUSYtransformation}) leads to the conditions
%
\begin{eqnarray}
&&\Delta^{T}G + G^{T}\Delta = 0\,, \label{condition1}\\
&&\langle\Delta\bar{\chi}_{\pm}\,,\,\phi_{\pm}*\phi_{\pm}\rangle
 + \langle\Delta\phi_{\pm}\,,\,\phi_{\pm}*\bar{\chi}_{\pm}\rangle
  + \langle\Delta\phi_{\pm}\,,\,\bar{\chi}_{\pm}*\phi_{\pm}\rangle = 0\,.
\label{condition2}
\end{eqnarray}
%
The first condition (\ref{condition1}) merely implies that
$\Delta^{T}G$ is antisymmetric, i.e. $(\Delta^{T}G)^{T} = - \Delta^{T}G$.
($\Delta^{T}$ denotes the transpose of the matrix $\Delta$.)
The second condition (\ref{condition2}) is nontrivial but
it is just the cyclic Leibniz rule (\ref{CLR}).
Thus, we have succeeded to obtain a lattice model invariant
under the $N=2$ nilpotent supersymmetry transformations 
(\ref{SUSYtransformation}) with the cyclic Leibniz rule.

%
\section{Holomorphy and non-renormalization theorem in the continuum}
%
%
An astonishing property of supersymmetric theories, which is not
shared by other theories, is the non-renormalization theorem.
To explain an essence of the theorem, let us consider, as an example, the 
four-dimensional Wess-Zumino model in the superspace formulation
\cite{Wess-Bagger}
%
\begin{eqnarray}
S_{WZ} = \int d^{4}x 
          \Big\{ \int d^{2}\theta d^{2}\bar{\theta}\
                   \Phi^{\dagger}(\bar{\theta})\, \Phi(\theta)
                 + \int d^{2}\theta\, W(\Phi)
                 + \int d^{2}\bar{\theta}\, \overline{W}(\Phi^{\dagger})
          \Big\}\,,
\label{WZmodel}
\end{eqnarray}
%
where $\theta, \bar{\theta}$ are the Grassmann coordinates.
We note that the first term, which is called the $D$-term, is given 
in a full superspace integral, while the second (or the third) term,
which is called the $F$-term, is represented in a half of the full
superspace integral.
The $\Phi(\theta)$ ($\Phi^{\dagger}(\bar{\theta})$) denotes a chiral
(antichiral) superfield and $W(\Phi)$ is called the superpotential,
which depends only on the chiral superfield $\Phi$ but not $\Phi^{\dagger}$.
The non-renormalization theorem tells us that there is no quantum
correction to the superpotential $W(\Phi)$ in any finite order of 
perturbation theory \cite{nonrenormalization}.

A simple proof of the non-renormalization theorem was given by
Seiberg \cite{holomorphy}, where the holomorphic property of the 
superpotential plays a crucially important role in the proof.
The tree-level superpotential $W=W(\Phi;m,\lambda)$ depends on the
(complex) mass and coupling constant, $m$ and $\lambda$, as well as the chiral 
superfield $\Phi$.
An important fact  is that an effective superpotential $W^{\footnotesize \textrm{eff}}$,
which comes from perturbation theory, could depend on $\Phi^{\dagger}, 
m^{*}, \lambda^{*}$ but actually holomorphy forbids them to appear
in $W^{\footnotesize \textrm{eff}}(\Phi;m,\lambda)$.
This can be proved by using a trick of extending the mass $m$ and the coupling
constant $\lambda$ to the chiral superfields $m(\theta)$ and $\lambda(\theta)$
\cite{holomorphy}.
The holomorphic property with some global $U(1)$ symmetries is found to be enough to show
that $W^{\footnotesize \textrm{eff}}(\Phi;m,\lambda) = W(\Phi;m,\lambda)$
in perturbation theory, as was proved by Seiberg.

%
\section{Superfield formulation and non-renormalization theorem on a lattice}
%
%
In order to preserve the holomorphic property on a lattice, 
one may try to define chiral superfields.
However, we again face an obstacle:
A chiral superfield $\Phi$ could be defined on a lattice by a
chiral condition $\bar{D}\Phi=0$ with a lattice supercovariant
difference operator $\bar{D}$.
It then turns out that any products of \lq\lq chiral\rq\rq\
superfields will not be chiral any more on a lattice.
This is because $\bar{D}$ should contain a difference operator 
and any difference operator cannot satisfy the Leibniz rule 
on a lattice, as was mentioned in Section 2.

Although we cannot introduce chiral superfields on a lattice well-definedly,
our lattice action (\ref{SCQM}) can be expressed in an $N=2$ superfield
formulation such that
%
\begin{eqnarray}
S = S_{\footnotesize \textrm{type I}} + S_{\footnotesize \textrm{type II}}\,,
    \label{action}
\end{eqnarray}
%
where
%
\begin{eqnarray}
S_{\footnotesize \textrm{type I}}
  &\equiv& \int d\theta_{-}d\theta_{+}\ 
         \langle\Psi_{-}(\theta_{+},\theta_{-})\,,\,\Psi_{+}(\theta_{+},\theta_{-})
         \rangle\,,
         \label{typeI}\\
S_{\footnotesize \textrm{type II}}
  &\equiv& \int d\theta_{-}d\theta_{+}\ \big\{
         \theta_{-}\,W_{+}(\Psi_{+},\Phi_{+};m_{+},\lambda_{+})
         - \theta_{+}\,W_{-}(\Psi_{-},\Phi_{-};m_{-},\lambda_{-}) \big\}
         \label{typeII}
\end{eqnarray}
%
with
%
\begin{eqnarray}
W_{\pm}(\Psi_{\pm},\Phi_{\pm};m_{\pm},\lambda_{\pm})
  = m_{\pm}\langle\Psi_{\pm}\,,\,G\,\Phi_{\pm}\rangle
    + \lambda_{\pm}\langle\Psi_{\pm}\,,\,\Phi_{\pm} * \Phi_{\pm}\rangle\,.
         \label{superpotential}
\end{eqnarray}
%
We note that $S_{\footnotesize \textrm{type I}}$
($S_{\footnotesize \textrm{type II}}$) corresponds to the first term
(the second and third ones) in eq.(\ref{WZmodel}).
We may call $W_{\pm}$ superpotentials like $W$ in the Wess-Zumino model.
The $\Psi_{\pm}(\theta_{+},\theta_{-})$ and $\Phi_{\pm}(\theta_{\pm})$
are lattice superfields defined by
%
\begin{eqnarray}
%
\Psi_{\pm}(\theta_{+},\theta_{-})
  &\equiv& \chi_{\pm} + \theta_{\pm}F_{\pm}
           + \theta_{\mp}(\pm i\Delta\phi_{\pm}) 
           + \theta_{-}\theta_{+} (\pm i\Delta\bar{\chi}_{\pm})
           \nonumber\\
       &=& \chi_{\pm} + \theta_{\pm}F_{\pm}
           + \theta_{\mp}(\pm i\Delta \Phi_{\pm}(\theta_{\pm}))\,,
           \label{Psi}  \\
\Phi_{\pm}(\theta_{\pm})
  &\equiv& \phi_{\pm} \pm \theta_{\pm}\bar{\chi}_{\pm}\,.
           \label{Phi}
\end{eqnarray}
%

%
It will be an easy exercise to verify that the above action (\ref{action}) in the
superfield formulation gives the original action (\ref{SCQM})
in component fields.
In terms of the superfields, the supersymmetry transformations
(\ref{SUSYtransformation}) are simply replaced by
%
\begin{eqnarray}
\delta_{\pm}\, {\mathcal O} = \frac{\partial}{\partial \theta_{\pm}} {\mathcal O}
           \label{SUSYtransformation2}
\end{eqnarray}
%
for any superfield ${\mathcal O}$.
It should be emphasized that each term of the action in the superfield formulation is 
supersymmetric, and especially the invariance of the interaction terms
under the $\delta_{\pm}$-transformations is guaranteed by the relations
$\langle\,\Delta\Phi_{\pm}\,,\,\Phi_{\pm}*\Phi_{\pm}\,\rangle = 0$,
which is nothing but eq.(\ref{totalderivative}) derived from the
cyclic Leibniz rule (\ref{CLR}).

Surprisingly, the non-renormalization theorem is found to hold in our
lattice model without introducing chiral superfields.
A key property to prove it is given as follows:
The superpotential $W_{+}$ depends on the $+$ variables
($\Psi_{+},\Phi_{+},m_{+},\lambda_{+}$)
at tree level.
One may naively expect that a perturbative effective superpotential
$W^{\footnotesize \textrm{eff}}_{+}$ could depend on both $\pm$ variables
($\Psi_{\pm},\Phi_{\pm},m_{\pm},\lambda_{\pm}$).
However, it turns out that the $N=2$ nilpotent supersymmetry allows
$W^{\footnotesize \textrm{eff}}_{+}$ to depend only on the $+$ variables
$(\Psi_{+},\Phi_{+},m_{+},\lambda_{+})$ but not on the $-$ variables.
This can be shown by extending $m_{\pm}$ and $\lambda_{\pm}$ to the
constant superfields
%
\begin{eqnarray}
m_{\pm} &\longrightarrow& 
   m_{\pm}(\theta_{\pm}) \equiv m_{\pm} + \theta_{\pm}\,\eta_{\pm}\,,
   \nonumber\\
\lambda_{\pm} &\longrightarrow& 
   \lambda_{\pm}(\theta_{\pm}) \equiv \lambda_{\pm} + \theta_{\pm}\,\xi_{\pm}
   \label{constantsuperfield}   
\end{eqnarray}
%
with the supersymmetry transformation property 
(\ref{SUSYtransformation2}).\footnote{%
We can show that the type II action (\ref{typeII}) is still invariant under
the $\delta_{\pm}$-transformations with the extension (\ref{constantsuperfield}).
}
If any term $A_{+}$ in $W^{\footnotesize \textrm{eff}}_{+}$
contains some of the $-$ variables ($\Psi_{-},\Phi_{-}, m_{-}, \lambda_{-}$), 
then we find that $A_{+}$ can be expressed always into the form
$A_{+}(\Psi_{\pm}, \Phi_{\pm}; m_{\pm}, \lambda_{\pm})
  = \frac{\partial}{\partial \theta_{-}}
     K_{+}(\Psi_{\pm}, \Phi_{\pm}; m_{\pm}, \lambda_{\pm})$
for some function $K_{+}$.
This expression implies that 
$\int d\theta_{+}d\theta_{-}\,\theta_{-}A_{+} = 
\int d\theta_{+}d\theta_{-}\,K_{+}$
and hence it should be treated as the type I term but not the type II term.
This observation tells us that our lattice model can possess
the holomorphic property as in the continuum theory.
By using holomorphy as in the continuum, we can show that 
the non-renormalization theorem holds even in our lattice model.
The details of the proof will be reported in Ref.\cite{NRT}.

Before closing this section, let us discuss an extension of 
$S_{\footnotesize \textrm{type I}}$ and $S_{\footnotesize \textrm{type II}}$
given in eq.(\ref{typeI}) and eq.(\ref{typeII}).
The type I action (\ref{typeI}) can be extended as
$S_{\footnotesize \textrm{type I}}
  = \int d\theta_{+}d\theta_{-}\ K(\Psi_{\pm}, \Phi_{\pm})$.
It is easy to verify that $S_{\footnotesize \textrm{type I}}$ is
supersymmetric for any function $K$ of $\Psi_{\pm}$ and $\Phi_{\pm}$
with the transformation property (\ref{SUSYtransformation2}).
On the other hand, the generalization of the type II action
is found to be nontrivial.
We note that if $W_{\pm}$ have the forms 
$\frac{\partial}{\partial \theta_{\pm}}K_{\pm}$,
they should be regarded as the type I terms.
This is because they become the type I terms by partial integrations
with respect to $\frac{\partial}{\partial \theta_{\pm}}$.
A detailed analysis shows that the $N=2$ nilpotent supersymmetry
severely restricts the superpotentials $W_{\pm}$ in the nontrivial type II terms 
to be of the form
%
\begin{eqnarray}
W_{\pm}(\Psi_{\pm},\Phi_{\pm})
 = \sum_{k} \lambda_{\pm}^{(k)}\,\langle\, \Psi_{\pm}\,,\,
    \underbrace{\Phi_{\pm}*\Phi_{\pm}*\cdots*\Phi_{\pm}}_{k-1}\,\,\rangle
         \label{generaltypeII}
\end{eqnarray}
%
with the cyclic Leibniz rule (\ref{k-CLR}).

%
\section{Summary}
%
%
In this paper, we have investigated a lattice model of supersymmetric complex
quantum mechanics, and proposed a cyclic Leibniz
rule as a modified Leibniz rule to preserve the $N=2$ 
nilpotent supersymmetry.
A notable property of our lattice model is the non-renormalization
of the potential terms in any finite order of perturbation theory.
Since the non-renormalization theorem is undoubtedly one of
the characteristic features of supersymmetric theories, our
results suggest that the cyclic Leibniz rule grasps some essence of
supersymmetry.
It would be of great importance to clarify the role of the cyclic
Leibniz rule in lattice supersymmetry.

In this paper, we have constructed the $N=2$ superfield formulation
for our lattice model.
It is interesting to rewrite the lattice action in the so-called
$Q$-exact form
%
\begin{eqnarray}
S = \delta_{+}\delta_{-}\,{\cal K}(\phi_{\pm}, F_{\pm}, \chi_{\pm}, \bar{\chi}_{\pm})
        + \delta_{+} {\cal W}_{+}(\phi_{+}, \chi_{+})
    + \delta_{-} {\cal W}_{-}(\phi_{-}, \chi_{-})\,,
    \label{Qexact}
\end{eqnarray}
%
where ${\cal K}$ is an arbitrary function of 
$\phi_{\pm}, F_{\pm}, \chi_{\pm}, \bar{\chi}_{\pm}$.
The ${\cal W}_{+}$ (${\cal W}_{-}$) depends only on $\phi_{+}$ and $\chi_{+}$
($\phi_{-}$ and $\chi_{-}$) and should satisfy the condition
%
\begin{eqnarray}
\delta_{\mp}{\cal W}_{\pm} = 0\,.
    \label{cohomology1}
\end{eqnarray}
%
If ${\cal W}_{\pm}$ were exact in a cohomological sense, ${\cal W}_{\pm}$ can
be written as
${\cal W}_{\pm} = \delta_{\mp}\,{\cal X}_{\pm}$.
Then, these terms should be classified as the first term in (\ref{Qexact})
but neither the second nor third term.
Therefore, ${\cal W}_{\pm}$ in the second and third terms of eq.(\ref{Qexact})
have to be closed but not exact with respect to $\delta_{\mp}$.
To find such terms is a problem of cohomology.
We can show that the cyclic Leibniz rule plays an important role in
classifying cohomologically nontrivial terms.
The details will be given in a forthcoming paper \cite{NRT}.

%
%
%


%
%
%
\end{document}